\documentclass{vgtc}                          %

\graphicspath{{figures/}{pictures/}{images/}{./}} %

\usepackage{times}                     %

\usepackage{tabu}                      %
\usepackage{booktabs}                  %
\usepackage{lipsum}                    %
\usepackage{mwe}                       %

\usepackage{mathptmx}                  %

\onlineid{0}

\vgtccategory{Research}

\vgtcinsertpkg

\title{A Scalable Approach to Clustering Embedding Projections}

\author{Donghao Ren\thanks{e-mail: donghao@apple.com}\\ %
        \scriptsize Apple %
\and Fred Hohman\thanks{e-mail: fredhohman@apple.com}\\ %
     \scriptsize Apple %
\and Dominik Moritz\thanks{e-mail: domoritz@apple.com}\\ %
     \scriptsize Apple
     }

\teaser{
  \centering
  \includegraphics[width=\linewidth,alt={}]{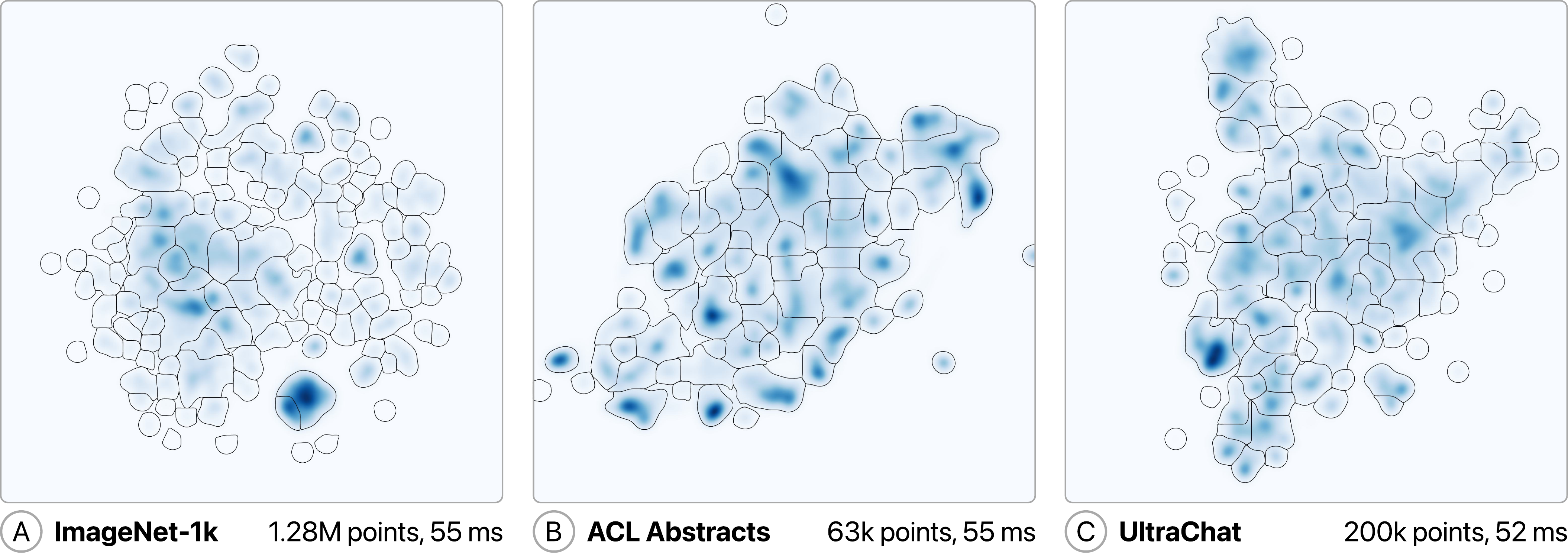}
  \caption{Three examples of clustering results over different datasets, including (A) ImageNet-1k~\cite{imagenet15russakovsky}, (B) ACL abstracts~\cite{wang2023wizmap}, and (C) UltraChat-200k from HuggingFace. The embeddings are projected with UMAP~\cite{mcinnes2020umap} using cosine distance. Time reported in milliseconds (ms) from clustering on a density map of $1000\times1000$ pixels. An open-source implementation of the algorithm is available at \url{https://github.com/apple/embedding-atlas}.}
  \label{fig:teaser}
}

\abstract{Interactive visualization of embedding projections is a useful technique for understanding data and evaluating machine learning models.
Labeling data within these visualizations is critical for interpretation, as labels provide an overview of the projection and guide user navigation.
However, most methods for producing labels require clustering the points, which can be computationally expensive as the number of points grows.
In this paper, we describe an efficient clustering approach using kernel density estimation in the projected 2D space instead of points.
This algorithm can produce high-quality cluster regions from a 2D density map in a few hundred milliseconds, orders of magnitude faster than current approaches.
We contribute the design of the algorithm, benchmarks, and applications that demonstrate the utility of the algorithm, including labeling and summarization.
} %

\keywords{Clustering, density data, embedding projections.}

\usepackage{xspace,xpunctuate}
\usepackage{algorithm2e}
\usepackage{amssymb}
\usepackage{enumitem}
\usepackage{hyperref}
\usepackage{tabularx}

\definecolor{gray}{RGB}{100, 100, 100}
\newcommand{\tableColor}[1]{{\textcolor{gray}{#1}\normalfont}}

\begin{document}

\definecolor{RoyalBlue}{HTML}{0071BC}
\newcommand{\todo}[1]{{\textcolor{RoyalBlue}{[#1]}\normalfont}}
\newcommand{\donghao}[1]{{\textcolor{RoyalBlue}{[#1 -DR]}\normalfont}}
\newcommand{\fred}[1]{{\textcolor{RoyalBlue}{[#1 -FH]}\normalfont}}

\newcommand{\ie}{{i.e.,}\xspace}
\newcommand{\eg}{{e.g.,}\xspace}
\newcommand{\ea}{{et~al\xperiod}\xspace}
\newcommand{\aka}{{a.k.a.}\xspace}
\newcommand{\etc}{{etc\xperiod}\xspace}
\newcommand{\etal}{{et al\xperiod}\xspace}

\firstsection{Introduction}

\maketitle

\textit{Embedding projection visualization} has proven to be a critical tool for machine learning (ML) development, from data exploration to model evaluation~\cite{hohman2018visual}.
Most ML models transform raw data (\eg images, text) into embedding vectors, but these vectors are hard to analyze as they have hundreds of dimensions without interpretable units.
To make sense of embeddings, practitioners use dimensionality reduction techniques (\eg PCA, t-SNE, or UMAP) to \textit{project} data from a high-dimensional space into a lower-dimensional one.
Since projection algorithms aim to preserve local neighborhoods and global structure from the high-dimensional embedding, the resulting low-dimensional projection is more tractable to visualize.

Embedding projection visualizations are typically represented as large scatterplots, where each point of data is represented by a point in the plane, and are useful for exploring a dataset and identifying clusters of similar points.
To make exploring a projection visualization useful, one needs to know what each point represents, but it can be tedious to inspect every point individually, especially as modern ML datasets grow.
Instead, some tools label clusters of points in the projection visualization to help give an overview of the projection.
Since the $x$ and $y$ coordinates for each point are generated from projection algorithms and are not predefined, the clusters and their labels are not known beforehand.
Generating clusters and their labels could be done manually, by inspecting and summarizing a visually salient group of data in the projection.
However, it has already been noted that this is intractable given the scale of data, which can easily contain millions of points and thousands of clusters.
Existing algorithms (\eg DBSCAN or Mean Shift) and popular toolkits (\eg scikit-learn) can generate clusters automatically, but they are computationally expensive since they operate on the point data.
In this paper, we take a practical approach and show that we can achieve much faster clustering speed by operating instead in the \emph{projected 2D space} using kernel density estimation.

Since it is standard practice to view and interpret clusters in the projected 2D space, we make three arguments for clustering and labelling in 2D with density rather than the high-dimensional space.
First, computing clusters and labels in the high-dimensional space results in clusters that spatially overlap once projected, which can be confusing to interpret and harder to visualize.
Multiple clusters might also exist at the same point once projected.
Operating on the 2D projected space will produce clusters that better align with what people perceive.
Second, computing clusters with point data is computationally expensive.
Instead of operating on the points directly, we can easily approximate the 2D projection with a density map, which does not scale as the number of points increases.
From the density map, we can develop efficient algorithms to produce cluster regions that mimic how people would identify clusters from the original data.
Lastly, clusters defined in the projection can be represented as 2D polygons, which is much easier to convert into database range queries to aggregate the underlying points, for example, to compute summary statistics or generate labels.

To help people interpret large projection visualizations, we take a practical approach and present an algorithm that is faster than point-based approaches for identifying clusters.
Given text data, we can also automatically label these clusters.
We demonstrate our approach clustering multiple ML datasets with more than one million points in around 100ms.

Our contributions include:

\begin{itemize}[noitemsep,topsep=0pt]
    \item \textbf{A fast and scalable algorithm} for clustering density data.
    \item \textbf{Algorithm complexity analysis and benchmarks} across three datasets, plus an interactive demo to illustrate its use in an embedding visualization tool.
    \item An open-source implementation of the algorithm, available at \url{https://github.com/apple/embedding-atlas}.
\end{itemize}

\section{Related Work}
\label{sec:related-work}

\subsection{Embedding Projection Visualization}

Embedding projection visualization is a popular approach for visualizing large ML datasets and internal model representations~\cite{hohman2018visual}.
Typical projection visualizations are represented as large-scale scatterplots, where each data point is plotted with $x$ and $y$ coordinates.
Since the axes and their units are defined by the projection algorithm, they may or may not be interpretable.
People instead use projections to visualize generated clusters of similar points.

ML datasets can easily contain millions of points.
Since each point is mapped to a single scatterplot point, projection visualizations often contain complex structure and suffer from overplotting~\cite{chen2014visual}.
To address these visual challenges~\cite{sarikaya2017scatterplots}, previous work has investigated alternative scatterplot designs, such as contour maps, hexbins, and design combinations like Splatterplots~\cite{mayorga2013splatterplots}.
More recently, literature has explored design spaces for aggregating and binning scatterplots to facilitate better data understanding~\cite{heimerl2018visual}.
While overplotting is a major concern in static scatterplot visualizations, most projection visualizations are typically included in visual analytic applications or have interactive support for zooming, panning, and filtering, rendering these challenges as negligible. 

Existing tools for projection visualization have continuously improved over the years, such as adding filters, interactions, and fast renders to scale to large datasets.
The Embedding Projector~\cite{smilkov2016embedding} is a good representative projection visualizer released to support visualizing datasets when ML toolkits, such as TensorFlow, were initially introduced.
WizMap is a more recent example that included fast searching for data points, contour maps to summarize data density, and automatic labelling~\cite{wang2023wizmap}.
Recent work has also investigated scaling these visualizations to millions of points in visual analytic systems, such as Nomic Atlas~\cite{nomic2022atlas}, and individual web components, such as Regl-Scatterplot~\cite{lekschas2023reglscatterplot}.
While these tools have advanced over the past decade, few have investigated and implemented clustering techniques to help people make sense of the now massive scale of these visualizations, and the few that do either have proprietary implementations or require a user to run this analysis themselves separate from the visualization.

\subsection{Density Data Clustering}

\begin{figure*}[tb]
 \centering
 \includegraphics[width=\textwidth]{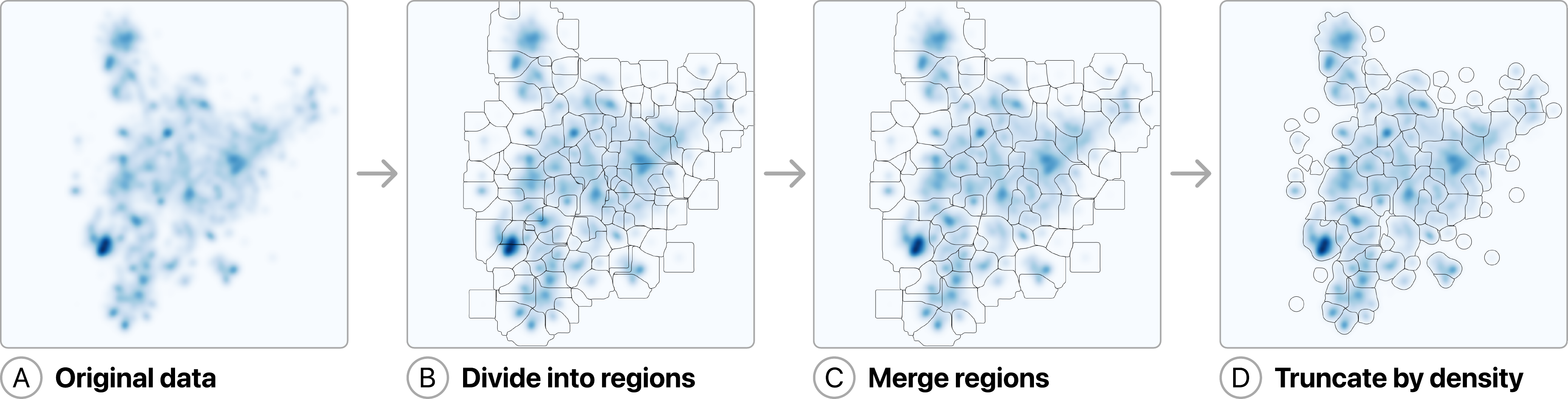}
 \caption{A visual explanation of the algorithm. Starting from a KDE of the (A) projected data, the algorithm first (B) divides the density data into regions using a disjoint set, (C) merges trivial regions with larger regions, and finally (D) truncates the regions by density levels into clusters.}
 \label{fig:algorithm}
\end{figure*}

Our approach falls into the broad category of density-based clustering techniques, where high-density, connected regions of data are grouped into clusters.
There are many existing density-based clustering techniques~\cite{musdholifah2013cluster}, such as DBSCAN~\cite{ester1996density}, GDBSCAN~\cite{sander1998density}, OPTICS~\cite{ankerst1999optics} and Mean Shift~\cite{comaniciu2002mean}.
These techniques are primarily designed for clustering high-dimensional data where it is not feasible to directly estimate the density function.
They often assume the input is a list of high-dimensional points and assign a cluster to each individual point.
However, in the case of clustering embedding projections in 2D space, we argue that it can be preferable to perform a kernel density estimation (KDE) first, and then cluster the density estimation instead of the full list of 2D points.
We can efficiently obtain the KDE through binning and approximation techniques~\cite{heer2021fast}; and when given the KDE, the runtime of our clustering algorithm depends only on the size of the density map and its complexity (\eg the number of resulting clusters). We extend the approach---proposed in Cytosplore~\cite{hollt2016cytosplore}---for generating the initial regions ((B) in \autoref{fig:algorithm}) with additional steps ((C) and (D)) to improve the quality of the resulting clusters.
An additional benefit of our approach is that it produces cluster regions as polygons instead of assigning a cluster to each individual point.
This makes it easier to post-process the clusters, such as displaying cluster regions or allowing a user to select a cluster by clicking.

In designing our approach, we took inspiration from level-set clustering algorithms~\cite{wang2009novel}.
The contour lines of a density map (\ie closed lines of equidensity locations) form a hierarchy where each lower-density contour contains a sub-tree of higher-density contours.
We can extract clusters from this hierarchy following certain criteria.
Most level-set clustering algorithms and analysis focus on high-dimensional data and generalizability~\cite{steinwart2011adaptive}.
In this paper, we take a practical approach in clustering 2D density estimations for visualization purposes.
In particular, we relax the requirement that a cluster must be derived from an equidensity contour, and allow clusters to be merged together to avoid superfluous clusters.

\section{Algorithm}
\label{sec:algorithm}

\RestyleAlgo{ruled}
\SetKwComment{Comment}{/* }{ */}
\SetKwFunction{Union}{Union}
\SetKwFunction{MakeSet}{MakeSet}
\SetKwFunction{UnionClusters}{UnionClusters}

\begin{algorithm}[!hbt]
\caption{2D density map clustering}\label{alg:algorithm}
\SetAlgoLined
\DontPrintSemicolon
\KwData{kernel density estimation as a 2D array $d(x, y)$}
\KwResult{cluster map $cmap(x, y)$}

\vspace{1em}
\Comment{Group pixels into initial clusters}
\ForEach{pixel $(x, y)$} {
    \MakeSet{$(x, y)$}\;
}
\ForEach{pixel $(x, y)$} {
    $(nx, ny) \gets$ neighbor pixel with maximum density\;
    \If{$d(nx, ny) \geq d(x, y)$} {
        \Union{$(x, y)$,$(nx, ny)$}\;
    }
}
\ForEach{distinct set of pixels $C$}{
    $id \gets$ new cluster id\;
    \ForEach{pixel $(x, y) \in C$}{
        $cmap(x, y) \gets id$\;
    }
}

\vspace{1em}
\Comment{Create a neighborhood graph of clusters}
$G \gets$ empty directed graph\;
\ForEach{pixel $(x, y)$ with cluster $c_1$} {
    \ForEach{neighboring pixel $(nx, ny)$ with cluster $c_2$} {
        \If{$c_1 \neq c_2$}{
            update $G$ at edge $c_1 \rightarrow c_2$ with the boundary pixel location $(x, y)$ and density $d(x, y)$\;
        }
    }
}

\vspace{1em}
\Comment{Union clusters}
\While{exists node pair in $G$ that satisfies the union criteria}{
    $(c_1, c_2) \gets$ the pair with highest priority\;
    \UnionClusters{$c_1$, $c_2$}\;
}

\vspace{1em}
\Comment{Truncate clusters by density threshold}
\ForEach{cluster $c$ in $G$}{
    \ForEach{pixel $(x, y)$ in $c$ where $d(x, y) < d_{min}(c)$}{
        $cmap(x, y) \gets \varnothing$\;
    }
}
\end{algorithm}

\autoref{alg:algorithm} presents pseudocode for the clustering algorithm.
The algorithm assumes a kernel density estimation is already computed on the original data set of projected $(x, y)$ coordinates.
This can be obtained efficiently using existing methods~\cite{heer2021fast}.

\paragraph{Generate Initial Clusters}
We first group pixels into initial clusters through hill-climbing.
Starting at a pixel $(x, y)$, in each iteration, move uphill to the neighboring pixel with the highest density.
This process finishes upon finding a local maximum.
Pixels can be clustered by which local maximum they land on.
We can efficiently implement this using a disjoint set data structure, where we union the set at each pixel with the set at its neighboring pixel with maximum density.
\autoref{fig:algorithm}B shows an example of the initial clusters.

\paragraph{Union Clusters}
The initial clusters may be fragmented since multiple local maxima could be close to one another, as shown in \autoref{fig:algorithm}B.
We attempt to union these clusters with neighboring clusters to improve readability.
The criteria for unioning two clusters is the following: if the location of the local maximum within a cluster is close to its boundary, union this cluster with the neighboring cluster sharing this boundary.
The idea is that if the local maximum is very close to a boundary, then it is likely this local maximum is merely slightly higher than the boundary and thus may not stand out much on its own.

Union is done by a greedy algorithm that prioritizes pairs that fit the criteria better (\eg by closeness to the boundary).
To efficiently implement this union process, we construct a cluster neighborhood graph $G$ that maintains information about the boundary between clusters and updates whenever two clusters union.
\autoref{fig:algorithm}C shows an example of merged cluster regions.

\paragraph{Truncate Clusters to Produce Clearer Cluster Boundaries}
The above steps will produce a cluster map where every pixel is assigned a cluster, as shown in \autoref{fig:algorithm}C.
However, due to the nature of kernel density estimation, there can be large swaths of regions with low density, and assigning a cluster to all these regions may not be ideal.
Therefore, in the final step, we truncate the cluster regions to a density lower bound, as seen in \autoref{fig:algorithm}D. The lower bound $d_{min}(c)$ for a given cluster $c$ is set to the maximum density in cluster $c$ times a constant factor ($0.1$ is used in this paper).

\paragraph{Post-processing}
The algorithm produces a cluster map that stores a cluster \texttt{id} at each pixel.
Depending on the use case, we can run a standard tracing method such as Suzuki's algorithm~\cite{suzuki1985topological} to convert the cluster map into boundary polygons.
These boundaries can be useful for displaying the cluster regions or querying the data for summarization (\eg \autoref{fig:labeling}).

\paragraph{Complexity}
The time complexity of the algorithm is $O(N \alpha(N) + M \log{M})$, where $N$ is the number of pixels in the density map ($N = width \times height$), and $M$ is the number of pairs of initial clusters that are neighbors.
We have observed that in a typical visualizations, $M$ is usually far less than $N$, so the runtime of the algorithm is dominated by the number of pixels $N$.
Note this algorithm scales with the size of the visualization and number of clusters, and is invariant to the number of data points.

\paragraph{Implementation}
We implemented this algorithm in Rust, and compiled it into WebAssembly for usage in Web environments.

\section{Evaluation}
\label{sec:results}

We evaluate the algorithm's runtime, clustering quality, and utility.
We compare our algorithm with a popular clustering library called supercluster\footnote{\url{https://github.com/mapbox/supercluster}} from Mapbox\footnote{\url{https://www.mapbox.com}}.
This library runs a modified version of DBSCAN, and it is one of the fastest libraries available for clustering 2D points. It is used in Mapbox for creating clusters of geographical points for summarization. We believe this is close to our use case in clustering 2D projections of embeddings.
We also experiment with a few notable algorithms including DBSCAN, HDBSCAN, OPTICS, and Mean Shift, with the implementations from the well-known Python package scikit-learn. We find that even for the small ACL Abstracts dataset with 63k points, these implementations are much slower than the supercluster library and our algorithm (1s for DBSCAN, 12s for HDBSCAN, and greater than 60s for OPTICS and Mean Shift). Furthermore, we also observe that significant parameter tuning is required to obtain reasonable clusters, likely because the default settings are designed for high-dimensional data. Thus, we only report comparisons with supercluster, with three datasets of increasing size.

\paragraph{Runtime}
We measure the performance of our implementation on three datasets on a MacBook Pro with an Apple M1 Pro processor (10 cores, 8 performance and 2 efficiency), and 32GB RAM. Our algorithm is implemented in native Rust; supercluster was run in Node.js 21.7.0. In general, our algorithm completes around 55ms when the input density map is $1000 \times 1000$. \autoref{fig:teaser}) shows three examples.
Note that the number of points is irrelevant since we assume that the density map is pre-computed. \autoref{tab:time-comparison} shows a more detailed comparison of the three datasets, including the time consumed to compute the kernel density estimation from 2D points in the GPU using WebGL. We observe that even with KDE time combined, our approach scales better than supercluster, especially for larger datasets. In addition, the combined time varies at around 80--100ms, which means it might be usable even in an interactive application.

\begin{figure}[tb]
 \centering
 \includegraphics[width=\linewidth]{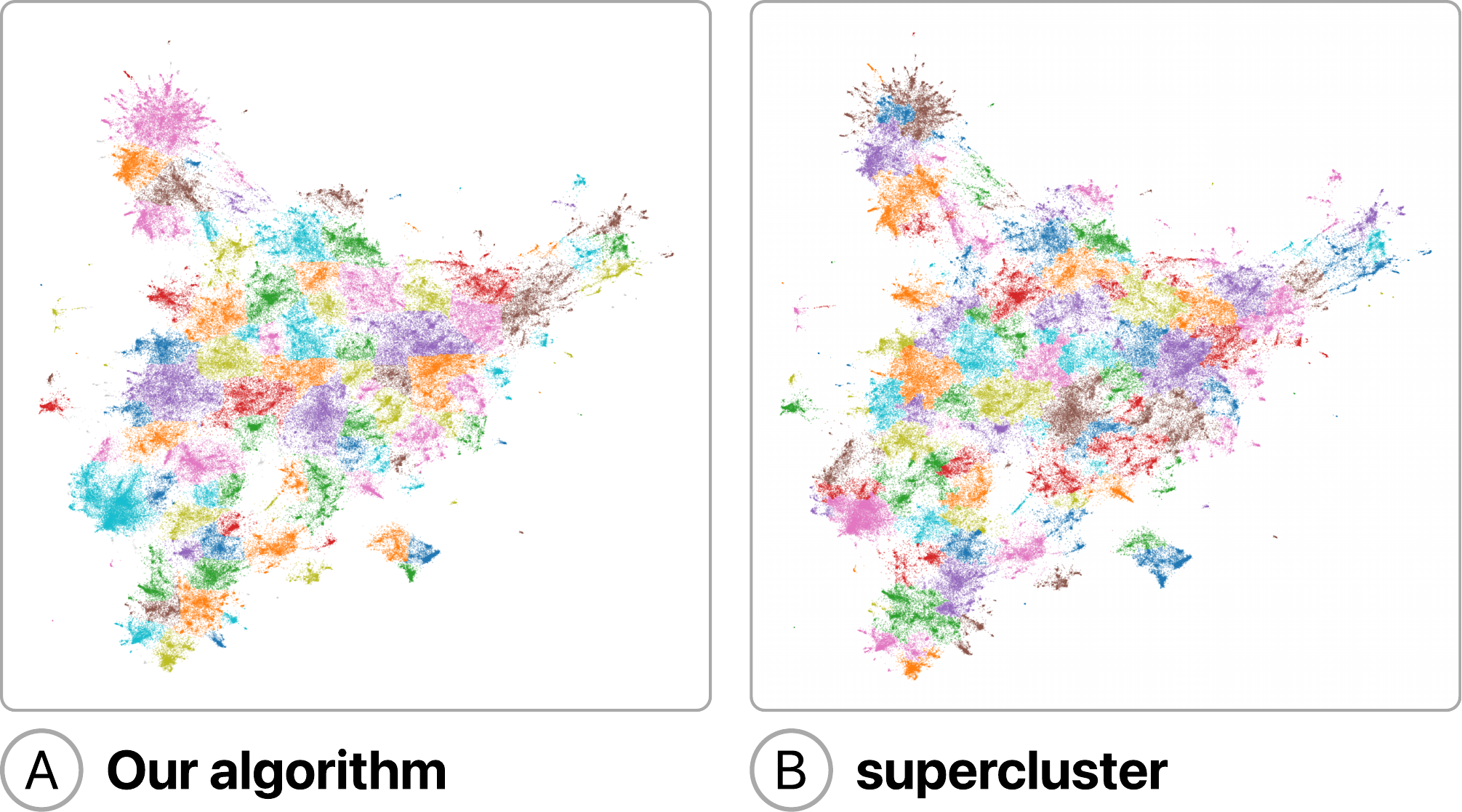}
 \caption{A comparison between (A) our algorithm and (B) supercluster, a popular library to cluster 2D points by Mapbox. For comparability, we assign a unique cluster \texttt{id} for points within each cluster region discovered by our algorithm, and use the cluster \texttt{id}s returned by supercluster. We also adjust the bandwidth and zoom level of the two algorithms to produce similar sized clusters. Since there are more clusters than colors that can be visually differentiated, we use a 10 color palette and ensure that adjacent clusters do not share the same color. Our algorithm takes 84ms to produce these clusters (time to compute KDE included), whereas supercluster takes 913ms to get the clusters and an additional 247ms to collect all the points from clusters.
 }
 \label{fig:comparison}
\end{figure}

\paragraph{Quality}
\autoref{fig:comparison} shows the cluster results from (A) our algorithm and (B) supercluster.
The two algorithms generate similar quality results, whereas our algorithm runs about 10x faster.

\begin{table}[tb]
  \caption{Clustering time comparison across datasets.}
  \label{tab:time-comparison}
  \begin{tabularx}{\linewidth}{X|rrrr}
    \toprule
      & \textbf{\# Points} & \textbf{Ours \tableColor{(+KDE)}} & \textbf{supercluster} \\
    \midrule
    ACL Abstracts   &   63K  & 55ms \tableColor{(+41ms)}  & 269ms   \\
    UltraChat-200k  &  200K  & 52ms \tableColor{(+32ms)}  & 913ms   \\
    ImageNet-1k     & 1.28M  & 55ms \tableColor{(+35ms)}  & 5611ms  \\
    \bottomrule
  \end{tabularx}
\end{table}

\begin{figure}[tb]
 \centering
 \includegraphics[width=\linewidth]{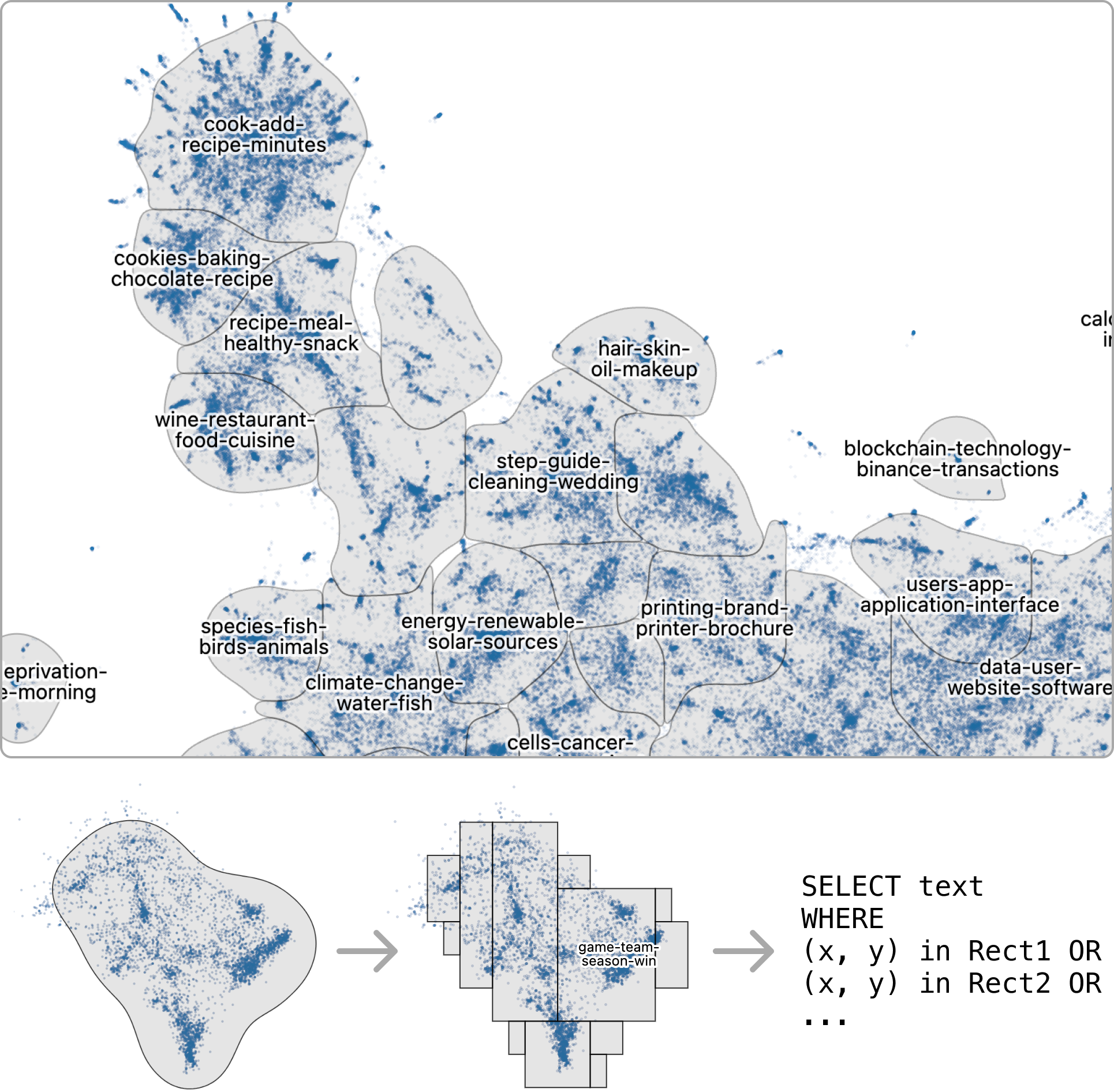}
 \caption{An example combining the clustering algorithm with automatic labeling. Top: We compute clusters for the UltraChat-200k dataset and label each cluster with a class-based TF-IDF method.
 Bottom: An illustration of one approach to query text data for generating labels. Starting with the cluster boundary polygon, we approximate the polygon with a set of axis-aligned rectangles, and then generate an SQL query with predicates testing for each rectangle. The \texttt{WHERE} clause in this query is then used to compute the TF metric for each word. The entire label generation process can be implemented as SQL queries.
 }
 \label{fig:labeling}
\end{figure}

\paragraph{Utility}
\autoref{fig:labeling} shows a projection of the UltraChat-200k dataset with labels generated from each cluster. For each cluster, we convert the cluster boundary into a set of non-overlapping rectangles and then translate them into SQL range queries. This allows us to produce c-TF-IDF-based labels directly using SQL queries. We built a web-based embedding viewer with this approach for labeling. It uses DuckDB WebAssembly to execute SQL queries. For the UltraChat-200k dataset with 200k points, it takes around 36s to \textit{generate labels} for 250 clusters (with two zoom levels) identified by our algorithm.

\section{Discussion}
\label{sec:discussion}

\paragraph{Benefits of Clustering from Density Maps}

First, in 2D, a density map can be efficiently estimated. There are a few approaches to do this. If the data can be loaded into RAM, we can bin the data in linear time, and then run an efficient density estimation algorithm~\cite{heer2021fast} on the bins. Both of these can be done on a GPU with custom shaders. For larger datasets stored in a database, we can have the database create the density bins, and then run the density estimation algorithm. Clustering from the density map allows us to leverage the existing state-of-the-art of density estimation, and therefore reduce the time consumed for clustering. In contrast, most existing libraries that cluster from points require a list of all points. This is harder to retrieve when the data is large.

Second, clustering from density maps creates cluster boundaries as polygons which can be used to generate database queries for summarizing points (\eg in~\autoref{fig:labeling}). With a point-based clustering algorithm, similar summarization would require a list of points be passed as a parameter to the query.

\paragraph{Limitations and Future Work}

The current algorithm produces a flat list of clusters. One may run the algorithm on multiple density maps with different bandwidths to create multiple levels of clusters for more granularity. However, these clusters do not have a hierarchical relationship. Developing an algorithm that can produce hierarchical clusters is compelling future work.

Many ML datasets have metadata, \eg one or more categorical class variable(s). Our algorithm currently only clusters a single density map, so it is unable to take such metadata into account. Another interesting research direction is to consider what the algorithm should do in the presence of categorical variable(s).

Since our algorithm runs on the 2D density map, it is limited to the projected space of an embedding projection. While clustering in 2D may produce results that better mimic how a human would visually draw clusters from such projections, it does not faithfully represent clusters in the original high-dimensional data.

\section{Conclusion}

In this paper we present a fast and scalable algorithm for clustering density data, analyze its complexity through examples and compare it to other popular scalable algorithms.

\acknowledgments{We thank our colleagues at Apple, including Yannick Assogba and Mary Beth Kery, for giving feedback on early drafts of this work.
}

\bibliographystyle{abbrv-doi}

\bibliography{main}
\end{document}